# Evidence for a quantum-spin-Hall phase in graphene decorated with $Bi_2Te_3$ nanoparticles

## Creating topological insulating graphene by heavy nanoparticle decoration with extremely small amount

K. Hatsuda[1], H. Mine[1], T. Nakamura[2], J. Li[3], R. Wu[3], S. Katsumoto[2], J. Haruyama[1,2*]

[1]Faculty of Science and Engineering, Aoyama Gakuin University, 5-10-1 Fuchinobe, Sagamihara, Kanagawa 252-5258, Japan
[2]Institute for Solid State Physics, The University of Tokyo, 5-1-5 Kashiwanoha, Kashiwa, Chiba 277-8581, Japan
[3]Department of Physics and Astronomy, University of California, Irvine CA 92697-4575, USA
*Corresponding author. E-mail: J-haru@ee.aoyama.ac.jp

**Realization of the quantum-spin-Hall effect in graphene devices has remained an outstanding challenge dating back to the inception of the field of topological insulators. Graphene's exceptionally weak spin-orbit coupling—stemming from carbon's low mass—poses the primary obstacle. We experimentally and theoretically study artificially enhanced spin-orbit coupling in graphene via random decoration with dilute $Bi_2Te_3$ nanoparticles. Remarkably, multi-terminal resistance measurements suggest the presence of helical edge states characteristic of a quantum-spin-Hall phase; the magnetic-field and temperature dependence of the resistance peaks, X-ray photoelectron spectra, scanning tunneling spectroscopy, and first-principles calculations further support this scenario. These observations highlight a pathway to spintronics and quantum-information applications in graphene-based quantum-spin-Hall platforms.**

## INTRODUCTION

Graphene played a key historical role in the development of topological insulators (*1,2*)—materials that exhibit an electrically inert interior yet form exotic metals at their boundary. In 2005, Kane and Mele predicted that coupling between the spin and orbital motion of electrons turns graphene into a 'quantum-spin-Hall (QSH)' insulator that hosts spin-filtered metallic edge states with inherent resilience from scattering (*3*). These novel edge states underlie tantalizing technological applications for low-power electronics, spintronics devices, and fault-tolerant quantum computing (*4-6*). Although graphene's intrinsic spin-orbit coupling is far too weak to produce an observable QSH phase in practice, numerous alternative platforms were subsequently discovered, including HgTe (*7-10*) and InAs/GaSb quantum wells (*11,12*), $WTe_2$ (*13-15*), bismuthene (*16*), and the layered compound $Bi_{14}Rh_3I_9$ (*17,18*).

Ease of fabrication, measurement, and manipulation of graphene nevertheless continues to strongly motivate efforts at fulfilling Kane and Mele's original vision. Might it be possible to externally boost graphene's spin-orbit coupling to stabilize a robust QSH phase, in turn opening an appealing pathway towards applications? Numerous theory works have pursued this line of attack via introduction of foreign atoms or via substrate engineering (*19-22*); these methods have been predicted to elevate the bulk band gap for the QSH phase by several orders of magnitude compared to that in pure graphene. Implementation of these proposals has, however, so far proven challenging despite recent efforts (*23-26*).

Here we present the first experimental evidence for the formation of an 'engineered' QSH phase in a graphene device. Specifically, we explore graphene decorated with dilute, randomly positioned

Bi$_2$Te$_3$ nanoparticles. These nanoparticles carry giant spin-orbit coupling via. tunneling current and can thus significantly modify graphene's electronic structure even at very low coverages (*26*); they can also be inserted into the graphene lattice in a minimally invasive way. As theoretically demonstrated on heavy adatom/graphene systems (*e.g.*, Os atom) in our previous study (*20*), topological bulk gaps open due to the hybridization between the graphene's $\pi$ states (Dirac state) and the spin-orbit-split $d_{xz}$ and $d_{yz}$ Os-atom orbitals in the case of Os-adatom. The Dirac state of graphene is rather delocalized and, thus, it can highly mediate interactions among Os adatoms over a large range even under small coverages. Qualitatively similar results are expected in the present system with small coverage ratios of heavy nanoparticles. Most strikingly, we perform non-local resistance measurements on multiple devices and find *quantitative* agreement with the response expected from dissipationless edge-state conduction in a QSH phase. The perpendicular magnetic-field and temperature dependence of the observed resistance peaks, scanning tunneling spectroscopy (STS), and first-principles simulations further corroborate this picture. Our results re-establish graphene as an experimentally promising QSH medium and spotlight many avenues of future exploration.

**RESULTS**

In the present experiments, monolayer graphene is grown by chemical vapor deposition on a SiO$_2$/Si substrate (area ~1cm$^2$) and formed into Hall-bar patterns by argon-gas etching with six or four branches to Ti/Au electrode terminals; see Figs. 1A-D and Fig. S1 from Supplementary Material 1 (SM1). Similar multi-terminal devices have been used to detect helical edge states in HgTe quantum wells (*8*). High quality of the monolayer graphene (i.e., low amounts of defects and contamination) is confirmed by Raman spectroscopy and X-ray photoelectron spectroscopy (XPS).

We further deposit Bi$_2$Te$_3$ nanoparticles with diameters of ~1-30 nm (Sigma Aldrich Inc.) onto the graphene surface following our previous nanoneedle method (SM2 and *26*). Specifically, for the present experiment, we repeatedly drop and then absorb a small acetone droplet containing the nanoparticles using the narrow tip of the needle (Saito Medical Instruments Inc.), which has an inner-pore diameter of ~50 μm, allowing precise control of the low nanoparticle density $D$ within the small graphene Hall bar (SM2 and video). Figure 1E presents an atomic-force-microscope image of a decorated sample with mean $D \sim 4 / 100^2$ nm$^2$ (~3% coverage ratio for the particles with ~10nm diameter), which we used for the present experiments. This coverage ratio may become somewhat higher when sub-nanoparticles formed by ultrasonication exist (SM2), although they cannot be confirmed in Fig. 1E. XPS spectra of the samples after the annealing at 400 °C (SM2), Figs. 1F-H, demonstrate a C*1s* orbital peak (~282 eV) arising from Bi-C coupling (Fig. 1F) and a Te*3d$_{5/2}$* orbital peak (~574 eV) arising from Te-C coupling (Fig. 1H). These peaks suggest clean, low-damage decoration with Bi$_2$Te$_3$ nanoparticles, and also indicate the hybridization required for enhancing graphene's spin-orbit coupling as mentioned above.

Figure 2 presents four-probe resistance measurements obtained in the six- and four-terminal (branch) devices from Figs. 1A and 1C. Current $I_{ij}$ flows from lead i to lead j and the voltage $V_{kl}$ is measured across contacts k and l, yielding a resistance $R_{ij,kl} = V_{kl}/I_{ij}$ that we monitor as a function of back-gate voltage $V_{bg}$. In a QSH phase, conduction is mediated by helical edge states that equilibrate at the contacts but are otherwise protected from elastic backscattering by time-reversal symmetry. Landauer-Buttiker formalism (*9*) then predicts $R_{ij,kl}$ values quantized to rational fractions of the resistance quantum $R_Q = h/e^2$, where h is Planck's constant and e is the electron charge. Quantized non-local resistances have indeed been previously reported in HgTe quantum wells as evidence for dissipationless helical edge transport (*9*). Figure 2A illustrates $R_{16,34}$ measured for undecorated graphene and with Bi$_2$Te$_3$ nanoparticles at coverage ~3%. As expected, without nanoparticles $R_{16,34}$ (measured at room temperature) essentially vanishes for all $V_{bg}$. Nanoparticle decoration, by contrast, yields an appreciable non-local resistance $R_{16,34}$ even at room temperature. Most interestingly, upon cooling down to $T = 1.5$ K we find an extended $V_{bg}$ window for which a maximum of $R_{16,34} \approx R_Q/6$ appears—in quantitative agreement with the value expected from helical-edge transport.

All other panels in Fig. 2 correspond to measurements on $Bi_2Te_3$-decorated graphene at $T = 1.5$ K. Figure 2B plots $R_{13,46}$, which peaks at $\sim 2R_Q/3$, while Fig. 2C shows $R_{14,23}$, which peaks at $\sim R_Q/2$. (The pronounced minimum in $R_{14,23}$ at $V_{bg} \sim 10$V possibly arises from a leakage current between the electrode pad 1 and back gate electrode through $p$-type silicon substrate). Figure 2D presents $R_{14,23}$ for a four-terminal $H$-shaped device at coverage $\sim 3\%$; this non-local resistance peaks at $\sim R_Q/4$. The peak values in Figs. 2B through 2D also agree quantitatively with the helical-edge-state picture. Four samples among the fabricated ten samples have demonstrated such $R$ peaks to date. However, the observed $R_{45,13}$ from our six-terminal device (Fig. 2E) significantly overshoots the resistance $R_Q/3$ expected from helical edge states. The origin of this discrepancy is presently unclear. We also detected nonlocal resistances in four other devices (not shown) that significantly undershoot the predicted quantized values—possibly indicating shorting of the current through the bulk of these samples.

Figure 3A illustrates examples of STS measurements on the decorated-graphene devices. No energy gaps are observed with $V_{bg}$, away from the resistance peaks in Fig. 2. Adjusting the $V_{bg}$ (e.g., 24V) to those coincident with resistance peaks in Fig. 2 yields qualitatively different behavior. For two spectra taken near to two different nanoparticles existing at bulk, clear spectral gaps are visible, ranging in magnitude from ~5-20 meV. This variation is consistent with simulations as explained later (Fig. 4 and SM3), which revel non-uniformity of these gaps as arising from variations in the nanoparticle size, chemical condition (*e.g.*, stoichiometry), and their chemical bonding with graphene Dirac states. By contrast, data taken at an edge point shows disappearance of such gaps. Note that these 'V-shaped' like edge spectra qualitatively resemble STS measurements in several other QSH candidates (*14,16,18*). Collectively, our STS experiments further support the emergence of an insulating bulk with gapless helical edge states driven by nanoparticle decoration.

Figure 3B shows the out-of-plane magnetic field ($B_\perp$) dependence of the conductance ($G$), corresponding to the inverse of the $R$ peak values of Figs. 2A (blue symbol) and 2D (red symbol) and the two off-$R$ peak values (green and pink symbols). The two $G$ values for the two $R$ peaks exponentially decrease with increasing $B_\perp$, whereas $G$ for the off-$R$ peak values decreases just slightly. These behaviors qualitatively agree with a QSH phase observed in monolayer $WTe_2$ (*15*) and support the attribution of the two $R$ peaks of Figs. 2A and 2D to the helical edge spins. A QSH phase emerges with a bulk energy gap but gapless helical edge states protected by time-reversal symmetry, in which opposite spin states form a Kramers doublet and counterpropagate. Only when a Fermi level is set to the Kramers point, $R$ values ($G$ values in Fig. 3B) can sufficiently reflect the band gap opening at the edges due to the Zeeman effect caused by applying $B_\perp$, resulting in the observed decrease in $G$ corresponding to the two $R$ peaks.

Almost the same slope values of the linear decreases in two $G$, for the six- and four-probe patterns, suggest that dephasing in metal electrodes is not much influenced by applying $B_\perp$ and only resolving in degeneracy in the Kramers doublet due to the Zeeman effect is caused over all the sample edges. This proves that the helical edge mode paths continuously exist along the sample edge, in spite of the presence of the multiple metal electrodes. The applied $B_\perp$ has almost no influence on the magnetization of the $Bi_2Te_3$ nanoparticles, because almost no magnetization is observed in the magnetization curve (*i.e.*, the amplitude of the diamagnetism observed around $B = 0$ is as small as $\sim 10^{-7}$ emu and the magnetism observed at high $B$ is smaller than $\sim 10^{-6}$ emu) even for graphene decorated with $Bi_2Te_3$ nanoparticles at a coverage as large as $\sim 10\%$ on application of $B_\perp$ (Fig. 3C) (SM2). The coverages of the samples used for the R measurements are much smaller (< 3%) and, hence, magnetization is almost absent.

The Zero-B temperature dependence of G (in Alenius plot format) corresponding to the inverse of the R peak values of Figs. 2A (blue symbol) and 2D (red symbol) is shown in Fig. 3D. The G values are constant at low temperatures, while they linearly increase above $T_c$ = 20 K and 25 K for the blue and red symbols, respectively, as temperature increases, following thermal activation relationships. The slope values of the linear increases in two G are slightly different with activation energies of ~20 meV and ~25 meV for the blue and red symbols, respectively, which approximately agree with the bulk gap values observed in the STS. Three possible origins are considered for these deviations from $(R_Q/6)^{-1}$ and $(R_Q/4)^{-1}$ at high temperatures: (1) the contribution of thermally activated carriers over the bulk gap, (2) spin fluctuation in helical edge spins, and (3) induced dephasing in metal electrodes. The agreement with the STS observation suggests that (1) is the dominant factor. Because the activation energy observed at B = 0 correspond to the bulk gap, this is consistent. Slightly lower $T_c$ (= 20 K) in the six-probe sample compared with $T_c$ (= 25 K) in the four-probe sample is also attributed to the smaller bulk gap.

**DISCUSSION**

To explain these experimental findings, we performed density functional theory (DFT) calculations using a large 7×7 graphene supercell containing a $Bi_2Te_3$ nanoparticle with 10 Bi and 15 Te atoms (the diameter of this cluster is about 1.2 nm). Since the atomic structure of the nanoparticle is unknown, we performed ab initio molecular dynamics (AIMD) simulations by baking it at 600K for 6 picoseconds and then cooling it down to 300 K in 5 picoseconds. The structures obtained through AIMD simulations were further relaxed at 0 K, with the inclusion of the van der Waals correction in DFT calculations. Figures 4A and 4B show side and top views of the optimized structure of the $Bi_2Te_3$ nanoparticle on graphene. The separation between the nanoparticle and graphene is about 3.4 Å, indicating weak interaction between them—contrary to single Bi atoms, which hybridize strongly with graphene. Moreover, a small corrugation appears in the graphene layer. From the charge-density difference ($\Delta\rho=\rho_{BT/Gr}- \rho_{BT}-\rho_{Gr}$), we see that Bi atoms donate electrons to graphene whereas Te atoms gain electrons from graphene.

Due to the weak van der Waals interaction, the composite system's band structure continues to exhibit Dirac cones (Fig. 4C). Electronic states of the $Bi_2Te_3$ nanoparticle reside rather far from the Fermi level and disperse weakly, indicating adequacy of the 7×7 supercell for minimizing direct interaction among adjacent nanoparticles. Significantly, the $Bi_2Te_3$ nanoparticles nevertheless yield a sizeable band gap $E_g \approx$ 6 meV at the Dirac point, which sits slightly away from the Fermi level due to the aforementioned charge transfer (Fig. 4D). To determine if the band gap is topologically nontrivial, we calculated the n-fields and $Z_2$ invariant from the Bloch functions (27,28) (Fig. 4E). By counting the positive and negative n-field numbers over the half of the torus—see Fig. 4E—one obtains a nontrivial $Z_2$ invariant. Test calculations with different k-point meshes consistently reproduce this result. Therefore, DFT predicts that $Bi_2Te_3$-nanoparticle-decorated graphene realizes a QSH phase, supporting our experimental observations.

To establish robustness of the topological state, we calculated the band structures and $Z_2$ invariants using an 8×8 supercell with $Bi_2Te_3$ nanoparticles containing an additional Bi or Te atom. The altered chemical stoichiometry of the nanoparticles shifts the Fermi level as shown in Fig. S3 (SM3). Furthermore, the reduction of $Bi_2Te_3$ coverage in the 8×8 supercell produces a band gap that is reduced (Fig. S2) yet remains topologically nontrivial in both cases. The band-gap magnitude and Fermi-level position thus appear to be tunable by adjusting the size, coverage, and stoichiometry of $Bi_2Te_3$ nanoparticles. These features are consistent with STS observations, which demonstrated non-uniform gaps depending on nanoparticles (Fig. 2F), and are clearly attractive for the development of graphene-based QSH devices.

**CONCLUSION**

In conclusion, multi-terminal $R$ measurements, those $B_\perp$ and temperature dependence, XPS, STS, and first-principles calculations provided strong evidence that helical edge states characteristic of a QSH phase emerge upon random decoration of dilute $Bi_2Te_3$ nanoparticles (as small as ~3%) into graphene via our nanoneedle method. On the contrary, edge states can arise even in topologically trivial systems (*29-31*), *e.g.*, due to band bending (*30*) and edge defects (*31*) as observed recently in ordinary graphene near the Dirac point. Although quantized edge resistances were not reported in those graphene studies, it may be important to ask whether our experiments are compatible with such a trivial scenario. Systematically exploring the effects of nanoparticle positions, sizes, and stoichiometry would help further clarification of the present QSH phase. It is highly expected that such experiments will introduce robust QSH phase to the present graphene and open the doors to innovative spintronic devices and quantum-information applications in graphene-based QSH platforms.

**MATERIALS AND METHODS**

Formation of graphene to small Hall-bar patterns with branches: CVD-grown monolayer graphene with $1cm^2$ area is formed into nine segments (see SM 1, Fig. S1A), including the two Hall-bar patterns with six or four branches in individual segments, by Ar gas etching (Fig. 1D and SM 1, Figs. S1B and S1C).

Nanoneedle decoration of $Bi_2Te_3$ nanoparticles: Before the $Bi_2Te_3$ nanoparticle decoration, an acetone solution containing the nanoparticles (Sigma Aldrich Inc.) is ultrasonicated for 3-5 hours by low power to obtain much smaller particle diameters (e.g., < 1nm) by crushing them, avoiding introduction of defects to nanoparticles. Then, low amount of acetone solution containing $Bi_2Te_3$ nanoparticles (e.g., 0.01mg/8ml) is dropped from the top end of the nanoneedle (Saito Medical Instruments Inc.) on the two neighboring Hall-bar patterns of graphene (SM 1, Fig. S1C). This droplet on graphene is then absorbed by the nanoneedle (see accompanying video). We repeat this dropping and absorbing 10-20 times, resulting in the controlled decoration of $Bi_2Te_3$ nanoparticles with very low density (< 5% coverage).

The obtained density and distribution of nanoparticles on the patterned graphene is sensitive to the density of nanoparticles contained in acetone solution, time for ultrasonication of the solution, position to absorb the solution by a nanoneedle from a container right after the ultrasonication, and the number of times to repeat the dropping and absorbing as mentioned above. Because the deposited nanoparticles are spread over a large area of the sample (*e.g.*, over the region for SM 1, Fig. S1B) depending on this optimized condition, a passivation film with an open window only on graphene surface (*e.g.*, SM 1, Fig. S1C) is needed. After the decoration, each sample is annealed at 400 °C for 10-15 minutes under a high vacuum ($10^{-6}$ Torr) for surface cleaning. These annealing conditions are the upper limit to keep quality of the nanoparticles. Indeed, nanoparticles annealed at 450 °C with the same time and vacuum degraded.

**SUPPLEMENTARY MATERIALS**

Supplementary material for this article is available at ・・・・.

**Acknowledgements**

JH thank J. Alicea, Y. Shimazaki, T. Yamamoto, S. Tarucha, T. Ando, S. Tang, Z-X. Shen, M. Dresselhaus, J. Shi, P. Jarillo-Herrero, A. H. Macdonald, and P. Kim for their technical contributions, fruitful discussions, and encouragement. **Funding:** The work at the Aoyama Gakuin University was partly supported by a grant for private universities and a Grant-in-Aid for Scientific Research (15K13277) awarded by MEXT. Work at the University of Tokyo was partly supported by Grant-in-Aid for Scientific Research (Grant No.26103003). DFT calculations at UCI were supported by DOE-BES (grant no. DE-FG02-05ER46237) and the computer simulations were supported by the National Energy Research Scientific Computing Center (NERSC). **Author contributions**: J. H., S. K., and R. W. conceived the ideas. K.H., H. M. and T.N. carried out the experiments and measurements. J. L. and R. W. performed the calculations. J. H. and R. W. wrote the manuscript. **Competing interests:** All authors declare that they have no competing interests. **Data and materials availability:** All data needed to evaluate the conclusions in the paper are present in the paper and/or the Supplementary Materials. Additional data related to this paper may be requested from the authors

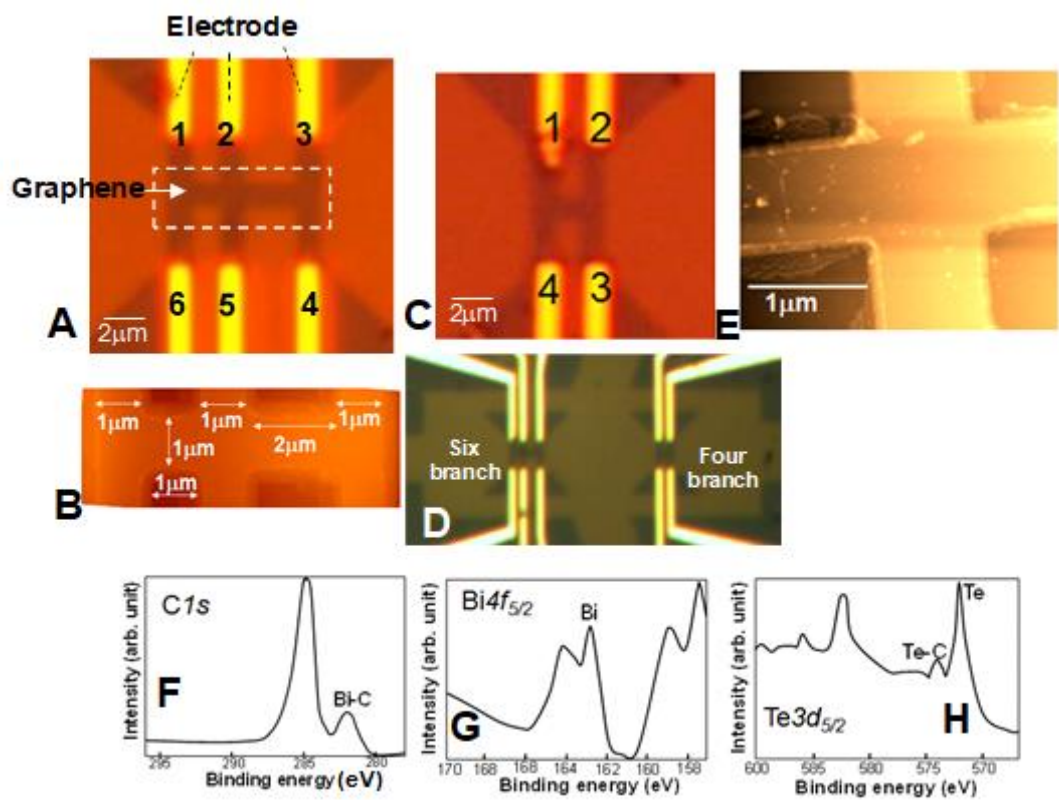

Fig. 1

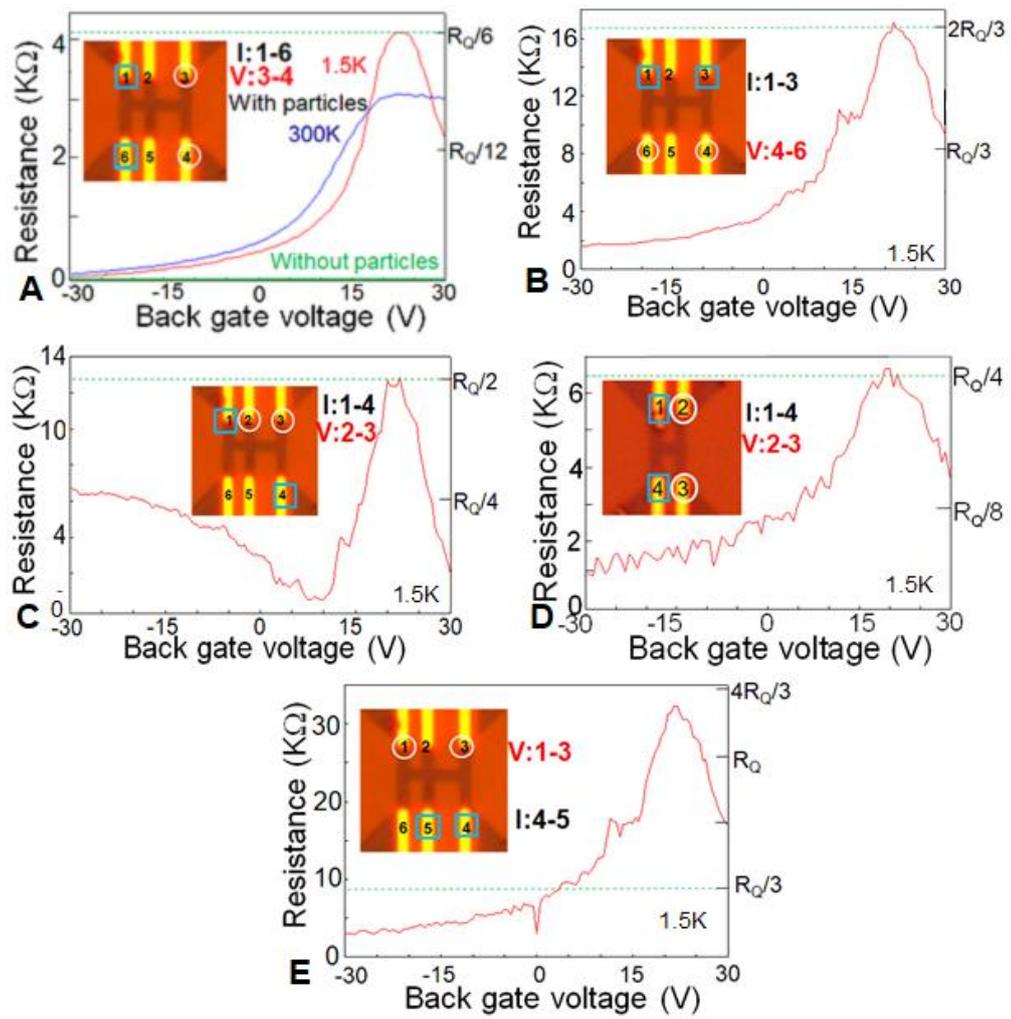

Fig. 2

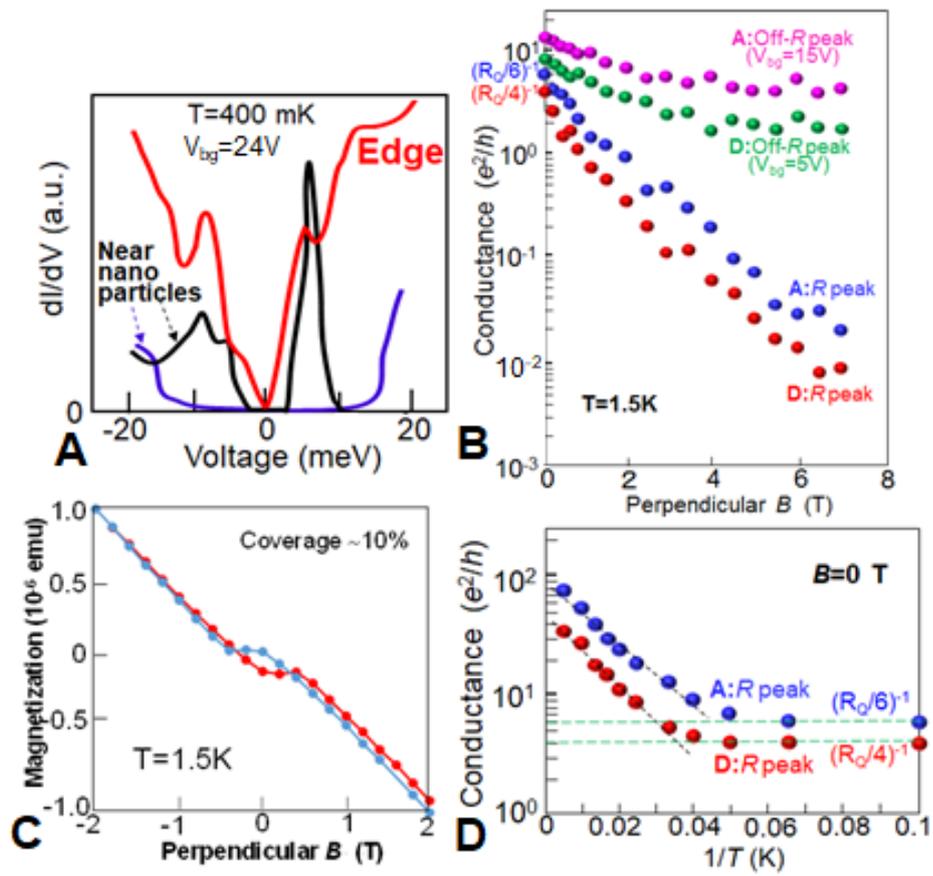

Fig. 3

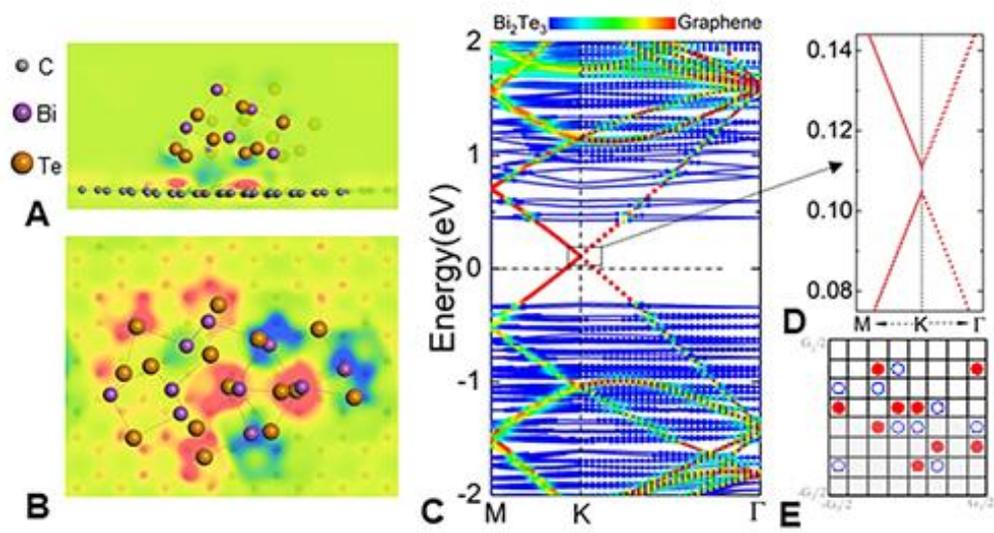

Fig. 4

**Figure captions**

**Fig. 1. Sample fabrication and characterization. (A-C)** Atomic-force-microscope (AFM) images of graphene Hall-bar devices used for resistance measurements of Fig. 2. Panel (B) shows an expansion of the dashed rectangle in (A). For (C), the channel and branch widths are ~1 μm. **(D)** Optical-microscope image of devices in (A) and (C), which are formed at neighboring position on the same segment of CVD-grown graphene (SM1). **(E)** AFM image of graphene decorated with $Bi_2Te_3$ nanoparticles at $D \sim 4 / 100^2$ nm$^2$ (~3% coverage ratio); the image is taken at the center of the six branches of (A). **(F-H)** XPS spectra of the samples.

**Fig. 2. Four-probe resistances versus back-gate voltage ($V_{bg}$) measured on the samples shown in Fig. 1.** Current flows between contacts indicated by squares, and voltage is measured across circled contacts; no contact resistances are subtracted. In (A), the green line corresponds to undecorated graphene. All other data corresponds to graphene with nanoparticles at ~3% coverage ratio. The green dashed line in each panel represents quantized resistances predicted for helical edge transport ($R_Q$ is the resistance quantum).

**Fig. 3. Reconfirmation of quantum spin Hall phase. (A)** STS spectra for a sample around $V_{bg}$ for $R$ peak, recorded at bulk locations near two different nanoparticles (~50 – 80 nm from particles), showing the maximum (~20 meV; blue line) and minimum (~5 meV; black line) gaps, and at an edge point (~50 nm from edges and ~200 nm away from nanoparticles), revealing the gap disappearance (red line). **(B)** Perpendicular magnetic-field ($B_\perp$) dependence of conductance ($G$) corresponding to the inverse of the $R$ peak values of Figs. 2A (blue symbol) and 2D (red symbol) and two off-$R$ peak values (pink and green symbols in Figs. 2A and 2D, respectively). **(C)** Magnetization curve for graphene decorated with $Bi_2Te_3$ nanoparticles at a coverage as large as ~10%, on application of $B_\perp$. **(D)** Zero-$B$ temperature dependence of conductance (in Alenius plot format) corresponding to the inverse of the $R$ peak values shown in Figs. 2A (blue symbol) and 2D (red symbol); the dashed lines serve as a guide to the eye.

**Fig. 4. Theoretical calculations. (A, B**) Side and top views of the atomic structure and charge-density difference of $Bi_2Te_3$/Graphene. Red and blue colors respectively indicate charge depletion and accumulation. **(C, D)** Band structure of $Bi_2Te_3$/Graphene. **(E)** The n-field configuration with red solid, blue hollow circles and blank boxes denoting n= -1, n= 1, and n=0, respectively. Summing the n-fields over half of the torus yields a nontrivial $Z_2$ invariant.